\documentclass[12pt,a4paper,final]{iopart}

\usepackage{iopams}  
\usepackage{graphicx}
\usepackage[breaklinks=true,colorlinks=true,linkcolor=blue,urlcolor=blue,citecolor=blue]{hyperref}
\usepackage[labelfont=bf]{caption}
\usepackage[usenames,dvipsnames,svgnames,table]{xcolor}
\usepackage{enumerate}
\usepackage{changepage}
\usepackage{multirow}	

\begin{document}

\title{Polarisation of physics on global courses}

\author{Allan L. Alinea}
\address{Department of Physics, Osaka University, Japan, 560 0043}
\ead{alinea@het.phys.sci.osaka-u.ac.jp}

\author{Wade Naylor}
\address{International College \& Department of Physics, Osaka University, Japan, 560 0043}
\ead{naylor@phys.sci.osaka-u.ac.jp}

\begin{abstract}
\noindent{Since October 2010, the Chemistry-Biology Combined Major Program (CBCMP), an international course taught in English at Osaka University, has been teaching small classes (no more than 20 in size). We present data from the Force Concept Inventory (FCI) given to first year classical mechanics students ($N=47$ students over three years) pre and post score, for a class that predominantly uses interactive engagement (IE), such as MasteringPhysics. Our findings show a $G$-factor improved score of about $\sim$ $0.18$, which is marginally about the average of a traditional based course. Furthermore, we analyse in detail a set of six questions from the FCI, involving the identification of forces acting on a body. We find that student answers tend to cluster about ``polarising choices"---a pair of choices containing the correct choice and a wrong choice with the latter corresponding to a superset of forces in the former. Our results are suggestive that students have a good idea of the right set of forces acting on a given system but the inclusion of extra force(s) brings about confusion; something that may be explained by misleading ontological categorisation of forces. In an appendix we also comment on possible correlations between the pre/post score and the level of English ability on entry to the course.}
\end{abstract}

\vspace{2pc}
\noindent{\it Keywords}: Physics Education; Interactive Engagement; Global Courses.
\maketitle

\section{Introduction}


\par In a typical calculus-based introductory mechanics course, a significant time is allotted in applying Newton's laws of motion. It follows that the skill in making Free Body Diagrams (FBD) (whether on paper or just inside one's head) is repetitively invoked throughout much of the course in solving mathematical and conceptual problems, and in trying to have a deeper understanding of the dynamics of several systems. Making a FBD is thus, one of the foundational skills a student is expected to acquire. Part of this foundational skill is the ability to identify the forces acting on a given body. If this very basic ability is not properly established, the student will have difficulty in gaining a deep understanding of Newtonian mechanics. For this reason, the need to look into this very simple matter of identifying forces cannot be overemphasised.

\par In the past couple of decades, much of the research involving students' understanding or the misconception of force and other related concepts in Newtonian mechanics focused on the ``cure'' in the form of interactive engagement (see for instance, \cite{Heller}, \cite{Hake}, \cite{Fagen}, \cite{Morote}, \cite{Brhane}). Although this is an equally important area of research, the subject itself of student misconception regarding force is far from being a closed-case. As we shall see, there are still avenues (eg., the existence of ``polarising choices'' (subsection \ref{IdentifyForce})) that need to be explored. New insights can be gained from this exploration through the administration of a concept inventory. In this work, we revisit the subject matter of understanding force with a special focus on the identification of forces through the use of the Force Concept Inventory (FCI).

\par The FCI developed by Hestenes, Wells, and Swackhamer in the early 1990s  is an instrument to assess student's understanding or misconception of force \cite{FCI}, \cite{Savi}. It consists of 30 multiple-choice questions each with five choices. After more than two decades of existence, there must be good reasons why it is still one of the most widely used concept inventories. Contained in the FCI are questions that suit our need to look into the subject of identifying force. This allows us to set aside the need to develop a new set of questions and focus on the analysis of the result of the FCI. The subjects of our testing are the students of the Chemistry-Biology Combined Major Program (CBCMP).

\section{Methodology}

\par The CBCMP is a program offered by Osaka University (Japan) that is geared for students coming from outside Japan.
The student composition is multiracial in nature with approximately 50\% coming from China  (including on and off the mainland), 25\% from Southeast Asia, and 25\% from other countries including Japan. Classes are conducted in English with the instructor meeting the class once a week for a period of 1.5 hours and a total of 20 hours in class (this is to be contrasted with a standard physics course which typically has 30/40 hours).

\par Having a limited number of class hours per week, traditional lectures can be both ineffective and impractical, so to deal with this shortage of hours we have employed MasteringPhysics \cite{MP}. In a typical class the students are given prelecture assignments \textcolor{black}{(10 questions each)} to read in advance before going to class and at every meet, the class starts with a set of prelecture conceptual questions some of which are discussed by the instructor. A short lecture then follows highlighting the main points of the scheduled subject of the meeting. For the remaining 60\% of the class hours, students are paired to discuss and answer the assigned problems (conceptual or quantitative) from MasteringPhysics. While answering the problems, the instructor and teaching assistant attend to student concerns about the problems they are solving. They are given one week to finish the assigned set of problems where students on average spend about \textcolor{black}{one to two} hours \textcolor{black}{doing their} homework.


\par We administered the FCI at the beginning and at the end of the course on calculus-based introductory mechanics. The data corresponding to a total of 47 students, were taken over three years from October 2011 until February 2014. The three-year data were subjected to statistical analysis mainly involving $ t $-test, graphical analysis, and calculation of the standard statistical parameters (eg., mean, standard deviation, etc).

\section{Results and Discussion}

\subsection{Result of the FCI as a Whole}
\begin{table}
\centering
    \begin{tabular}{|l|cccc|}
    \hline
    Group   & 2011 ($n=11$)  & 2012 ($n=18$)  & 2013 ($n=18$)  & Total ($N=47$) \\
    \hline
       & Mean (SD)    & Mean (SD)    & Mean (SD)    & Mean (SD)    \\
    Pretest & 17.1 (6.6) & 16.6 (7.4) & 14.7 (6.7) & 16.0 (6.9) \\
    Posttest & 20.2 (7.0) & 19.7 (6.3) & 16.4 (7.0) & 18.6 (6.8) \\
    Gain*    & 0.24         & 0.23        & 0.12      & 0.18        \\
    \hline
    \end{tabular}
    
    \caption{\it Result of the FCI for the three batches of entrants from 2011-2013. ${}^*$For the Normalised Gain standard deviation (SD) is not applicable.}
    \label{table_1}
\end{table}

\textbf{Table} \ref{table_1}\footnote{For a recent article on which kind of concept test to implement and how to interpret the data see \cite{Best}.} shows the result of the FCI for three batches of entrants to the CBCMP program. We would like to stress that given the number of students we have we found an overall difference in the means of the pre and post test scores in favor of the latter. To further verify this we used $R$ (statistical software) \cite{Rstat} to find the paired student $t$-test result. At the 95\% confidence level we found a $p$-value of $4.1\times 10^{-5}$ for 46 degrees of freedom with a mean difference of 8.6\%. The null hypothesis was not satisfied and suggests the results do lead to a positive increase in the post scores.


The $ t $-test only indicates whether there is a significant difference between the pre and post test scores. A better measure of student performance before and after instruction is provided by the \textit{normalised gain}, $ G $ \cite{Hake}. It is the difference between the pre and post test scores divided by the maximum possible score in the posttest relative to the score in the pretest. Symbolically,
\begin{equation}
	\hspace{5.0em} G = {\langle \% S_f\rangle  - \langle \% S_i \rangle \over 100 - \langle \% S_i \rangle },
\end{equation}
\noindent
where $\% S_f$ and $\% S_i$ are the final and initial scores respectively. \textbf{Table \ref{table_1}} shows the normalised gain for the three batches of students. On average, we find a result of $G = 0.18$. At first we found this result to be a little surprising, as we expected a higher gain in line with interactive engagement (IE) based courses \cite{Hake}, \cite{sokoloff}, \cite {coletta}. In the appendix, we look into the possible correlation of the normalised gain with the English language ability.



\subsection{Identifying Forces}
\label{IdentifyForce}
The relatively low normalised gain may be attributed to a variety of factors. The complexity of the FCI however, does not give us much freedom to investigate a lot of things. The FCI centring around the idea of force connects to so many other structures of human understanding regarding force that we are bound to focus only on a small manageable part. Beyond that, we can say a lot of things from logic alone, but scarcity of experimental results on our part would simply hinder us in drawing a sound conclusion. 

What we do in this work as already mentioned, is focus on the identification of forces. For this subject, we choose only segments of the FCI (labeled here as questions I to VI)\footnote{These correspond to questions 5, 11, 13, 18, 29, and 30.} that fall under the mentioned category. The hope is to see an obvious pattern on how students identify force. 

\begin{figure}[htb!]
    \centering
    \includegraphics[scale=0.85]{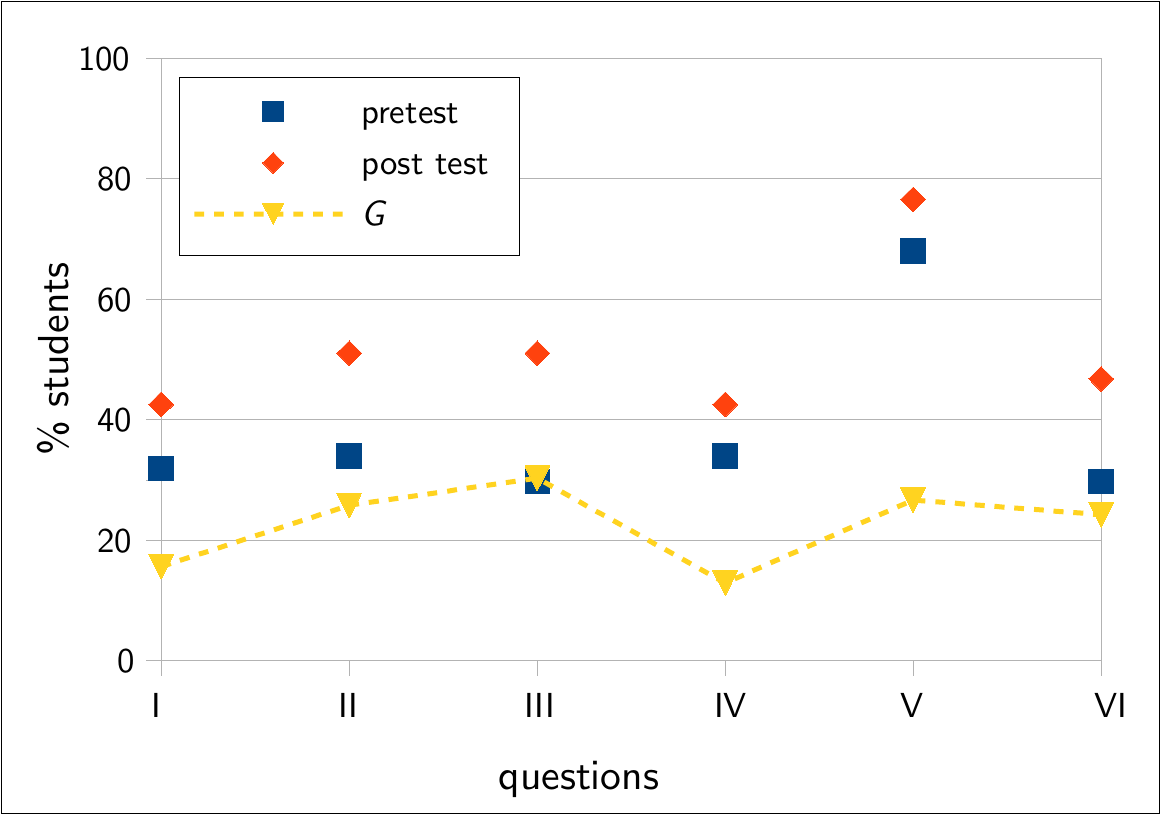}
    \caption{\it Normalised gain ($G$) (expressed in percent) for selected questions involving the identification of forces acting on a given body.}
    \label{fig_1}
\end{figure}

\par Let us now look into the pre and post test scores. \textbf{Figure \ref{fig_1}} shows the percent of students who got the correct answer for the six questions labelled I to VI involving identification of forces acting on a given body. The percent of students who got the correct answer in the posttest is higher than that of the pretest for all questions. However, similar to that of the overall FCI result presented in the previous subsection, the improvement is small. The \textit{normalised gain} expressed in percent ranges only from 13\% to 30\% which is classified as \textit{low G} in the work of \cite{Hake} (medium $G$ : 30\% \textless $G$ \textless 70\%, high G  : \textgreater 70\%).
\begin{figure}[htb!]
    \makebox[\textwidth][c]{\includegraphics[width=1.15\textwidth]{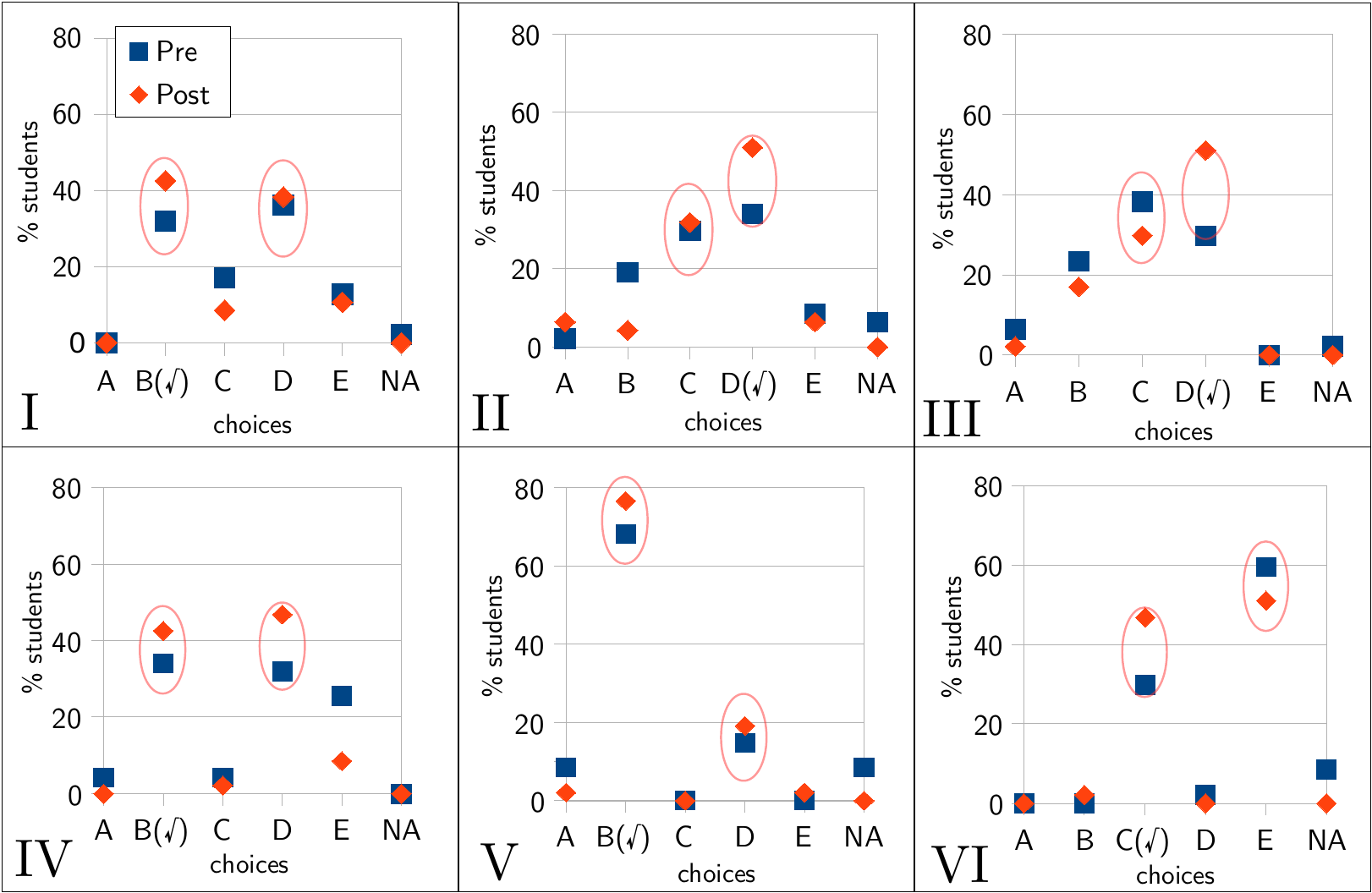}}
    \caption{\it Distribution of student answers for questions I to VI. The five choices are labeled A to E (NA means no answer). The letter of the correct answer is labeled with a check mark. The red circles indicate the top two choices ``polarising'' student answers.}
\label{fig_2}
\end{figure}
\begin{figure}[htb!]
    \centering
\scalebox{0.75}{\includegraphics{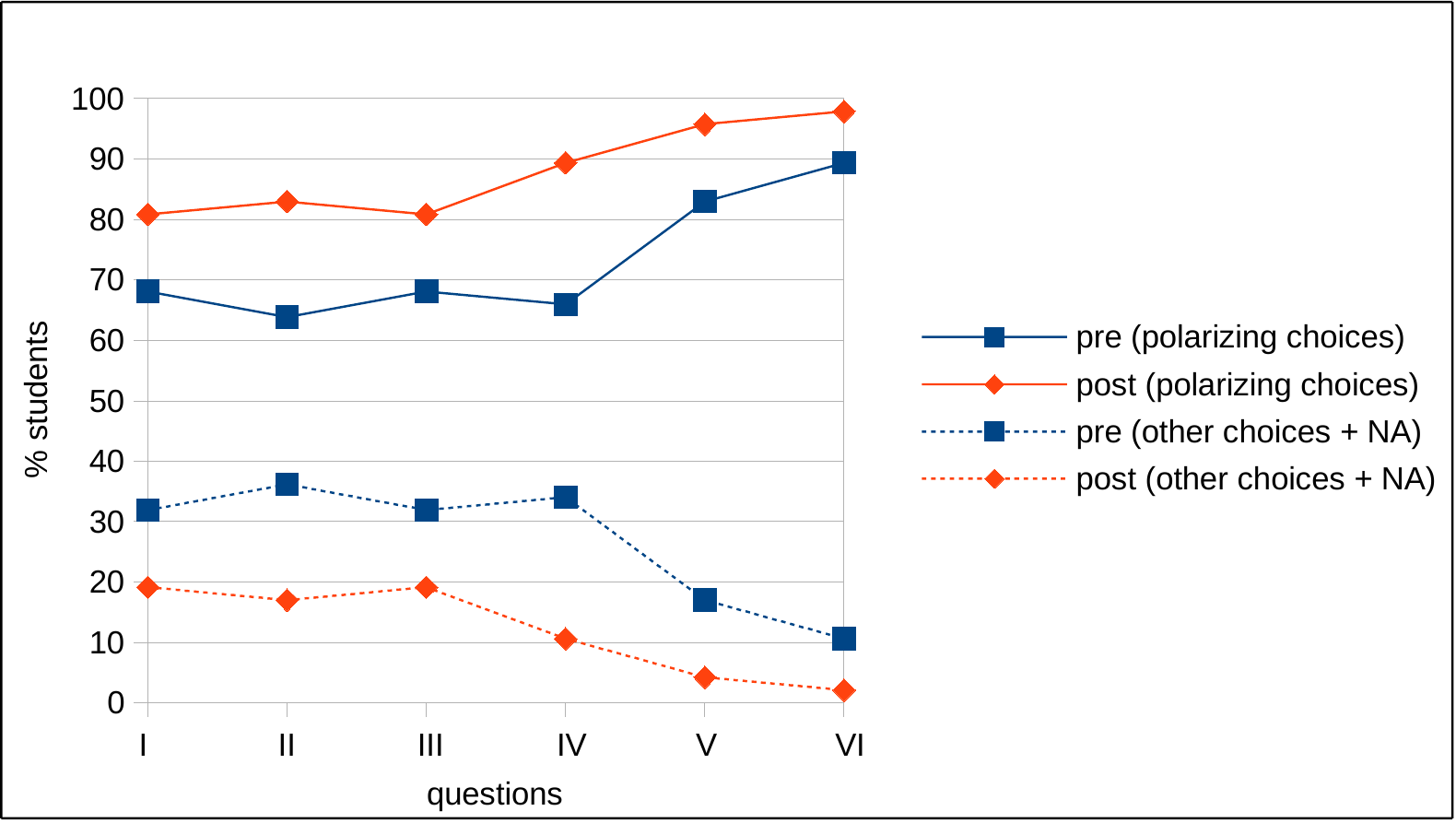}}
    \caption{\it Distribution of student answers to questions I to VI.}
\label{fig_3}
\end{figure}

\par The seemingly uninteresting result above prompted us to look closely on how the students answered the six questions. \textbf{Figure \ref{fig_2}} shows the result of the item analysis for each question. The number of students who answered each letter for a given question is counted and the result is plotted in the figure. Careful analysis of the results indicate that 
\bigskip
\begin{adjustwidth}{45pt}{}
	\begin{enumerate}[(a)]
		\item
		there are two dominant answers for both the pretest and the posttest,
		\item
		the pair of dominant answers is the same for both the pretest and the post test for each question,
		\item
		the pair of dominant answers contain the correct answer.
	\end{enumerate}
\end{adjustwidth}
\bigskip
By a ``pair of dominant answers,'' we mean a pair of choices with the highest frequency. Consider for instance, question I. The pair of dominant answers are choices B and D. For the pretest, the share of the pair of dominant answers is 68\% and it rises to 71\% in the posttest. Such a large share of the pair of dominant answers justifies our use of the term ``dominant''. \textbf{Figure \ref{fig_3}} shows the wide gap between the percent of students who answered the pair of dominant answers and those who answered the remaining choices for both the pretest and posttest. The percent of students who answered the dominant answers ranges from 64\% to 89\% for the pretest and from 81\% to 98\% for the posttest.

\par The curious pattern that we found out above regarding the existence of a pair of dominant answers raises the following concerns:
\bigskip
\begin{adjustwidth}{45pt}{}
	\begin{enumerate}[(a)]
		\item
		Why is there a pair of dominant answers?
		\item
		What is the nature of this pair of dominant answers?
		\item
		Why do students choose the pair of dominant answers?
	\end{enumerate}
\end{adjustwidth}
\bigskip
Observation of the plots in \textbf{Fig. \ref{fig_2}} in connection to the six FCI questions tells us that the pair of dominant answers always contain the correct choice; symbolically, it is (CC, X) where CC is the correct choice while X is the other choice. Furthermore, both CC and X contains the right set of forces acting on a given body \textit{but} X contains one or more extra forces that do/does not legitimately act on a given body. In other words, X is a superset of CC. The set of extra ``fictitious'' force(s) in X confuse(s) the student causing a sort of ``polarisation''; that is, in analogy to charge polarisation, the majority of student answers are divided into two namely, choices CC and X. For this reason, the more appropriate term for CC and X seems to be \textit{polarising choices} and we are going to use the term from hereon.

\par The existence of polarising choices brings the idea of misleading ontological categorisation \cite{JS}. When facing a problem requiring the identification of forces acting on a given body, several forces may come to mind. When there are only few forces acting on a given body, it is possible to have a good idea that includes all the valid forces plus the extra forces that are wrong. Whereas the matter of picking up the valid forces and putting it in the category of valid forces may be easily done after a careful analysis, the matter of categorizing the extra forces as correctly acting or not can be quite problematic. 

\begin{figure}[htb!]
    \centering
    \includegraphics[scale=0.35]{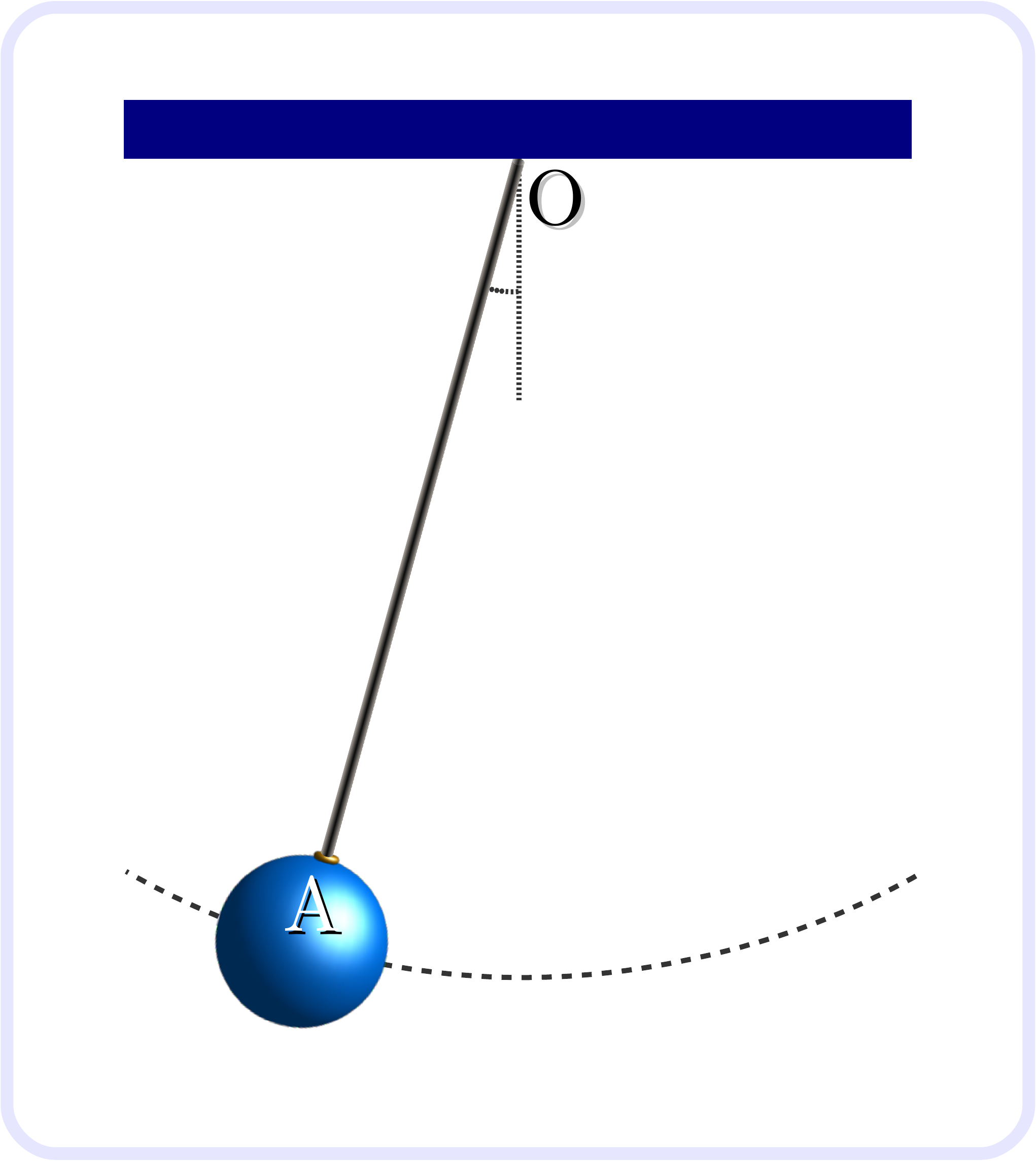}
    \caption{\it A typical question used to test conceptual ability in identifying forces. Consider a pendulum attached at $O$ swinging from a previous higher point to $A$. The students have to distinguish what are ``real" and ``fictitious" forces from: (i) the downward force of gravity, (ii) the tensional force exerted by the string from $A$ to $O$ and (iii) the force in the direction of the pendulum's motion.}
    \label{Ex_Sol}
\end{figure}

\par The idea of polarising choices can be made clearer through an example. Consider a snapshot of a swinging pendulum shown in \textbf{Fig. \ref{Ex_Sol}}. A student may be asked to identify the forces acting on the pendulum bob with the following choices: 
\bigskip
\begin{adjustwidth}{45pt}{}
    \begin{enumerate}[(i)]
        \item downward force of gravity
        \item the tensional force exerted by the string from $A$ to $O$
        \item the force in the direction of the pendulum's motion
    \end{enumerate}
    
    \medskip
    \begin{enumerate}[A.]
        \item (i) only
        \item (ii) only
        \item (iii) only
        \item (i) and (ii) only
        \item (i), (ii), and (iii)
    \end{enumerate}
\end{adjustwidth}
\bigskip

\noindent
The correct answer is letter D corresponding to forces in (i) and (ii). It is worth noting that choice E offers the same set of forces as that of choice D but with an additional ``fictitious'' force given by (iii). Choices D and E are our polarising choices for this question. An average student may be able to easily discard choices A, B, and C but may find choices D and E rather confusing. As we have seen in the results above, the inclusion of extra ``fictitious'' force for questions of this type can cause a serious divide in the answers of students. 

\par One curious thing that caught our attention regarding this matter of polarisation is that most students did well in their homework (average 88\%). The immediate question is ``How does the system of MasteringPhysics guide the students to the correct answer?'' MasteringPhysics \cite{MP} has a ``corrective measure'' by way of hints and trial-and-error. When it comes to answering problems requiring the identification of forces, the existence of hints together with the limited number of tries the students are allowed to make, narrows down the number of possible pathways to the correct answer in their homework.  Although we may need more data, our preliminary study here suggests that the ``corrective measure'' of the system of MasteringPhysics may not be able to correct deeply ingrained student misconceptions/preconceptions about force. However, we are not ignoring the possibility that we might also be able to use such a system to combat the misconception/preconception of extra force(s).

\section{Summary and Concluding Remarks}

\par In this article we have discussed several points involving (a) the notion of polarisation as it arises in answering questions about the identification of forces acting on a body, (b) the FCI as an instrument to assess student understanding of forces, (c) the interactive web-based homework and tutorial software MasteringPhysics, and (d) the English language ability as it relates to student learning (see the appendix). 
 
\par We looked at six particular questions on the FCI to establish if there were any patterns in answering questions. We came across a ``polarisation'' where student answers tend to cluster about two competing choices with the wrong choice being a superset of the right choice. Our results are suggestive that students have a good idea of the right set of forces acting on a given system but the inclusion of extra force(s) brings about confusion; something that may be explained by misleading ontological categorisation of forces. Progressive hints and limited number of tries offered by interactive web-based Physics education software in answering problems  involving proper identification of forces, could help students ``get to" the correct answer. This however, may not be enough to significantly alter misleading preconceptions or misconceptions of students.

\par In the appendix, we raised the issue of English ability and performance on the FCI (more generally any science-based test) and appeared to find \textit{no} correlation, most likely arising from the requirement of a TOEFL score above $\sim 80$. In Japan, at least there appears to be no minimum reading or writing score required 
and the bottom panel of \textbf{Figure \ref{fig_0}}  clearly appears to show a correlation (as might the Total TOEFL score if any range of students were accepted: TOEFL $0\sim 120$). This certainly warrants the implementation of a minimum reading and writing score on TOEFL, and similar tests, for entrance to global courses in Japan. 

\par What we have found out in this study open new avenues for future investigation. Firstly, with regard to the idea of polarising choices, we have so far only scratched the surface of answering intriguing questions such as (a) how students ontologically categorise forces acting on a given body, (b) why polarising choices exist (it may be constructed by design, such as via the use of ``distractors" \cite{Best}, but \textit{only} experiment can tell us whether there really is a resulting polarisation) and how such an idea relates to our way of thinking, (c) how we can ``cure'' misconceptions/preconceptions related to polarising choices and polarisation, and (d) whether polarising choices and the resulting polarisation can also be induced/found for other physics concepts aside from force. Secondly, the issue of English language ability-FCI for international courses is far from being settled. Although intuitively, there should be a positive-correlation between the two, questions on how far English ability is a factor on learning physics concepts for a multiracial class are something worthy of further investigation.

\section*{Acknowledgements}

We would like to thank current and past students on the Chemistry-Biology Combined Major Program (CBCMP) for creating stimulating interactive engagement classes.

\appendix
\setcounter{section}{1}
\section*{Appendix: English Language vs. Reading Ability}

\par
\begin{figure}[htb!]
    \centering
    \scalebox{0.6}{\includegraphics{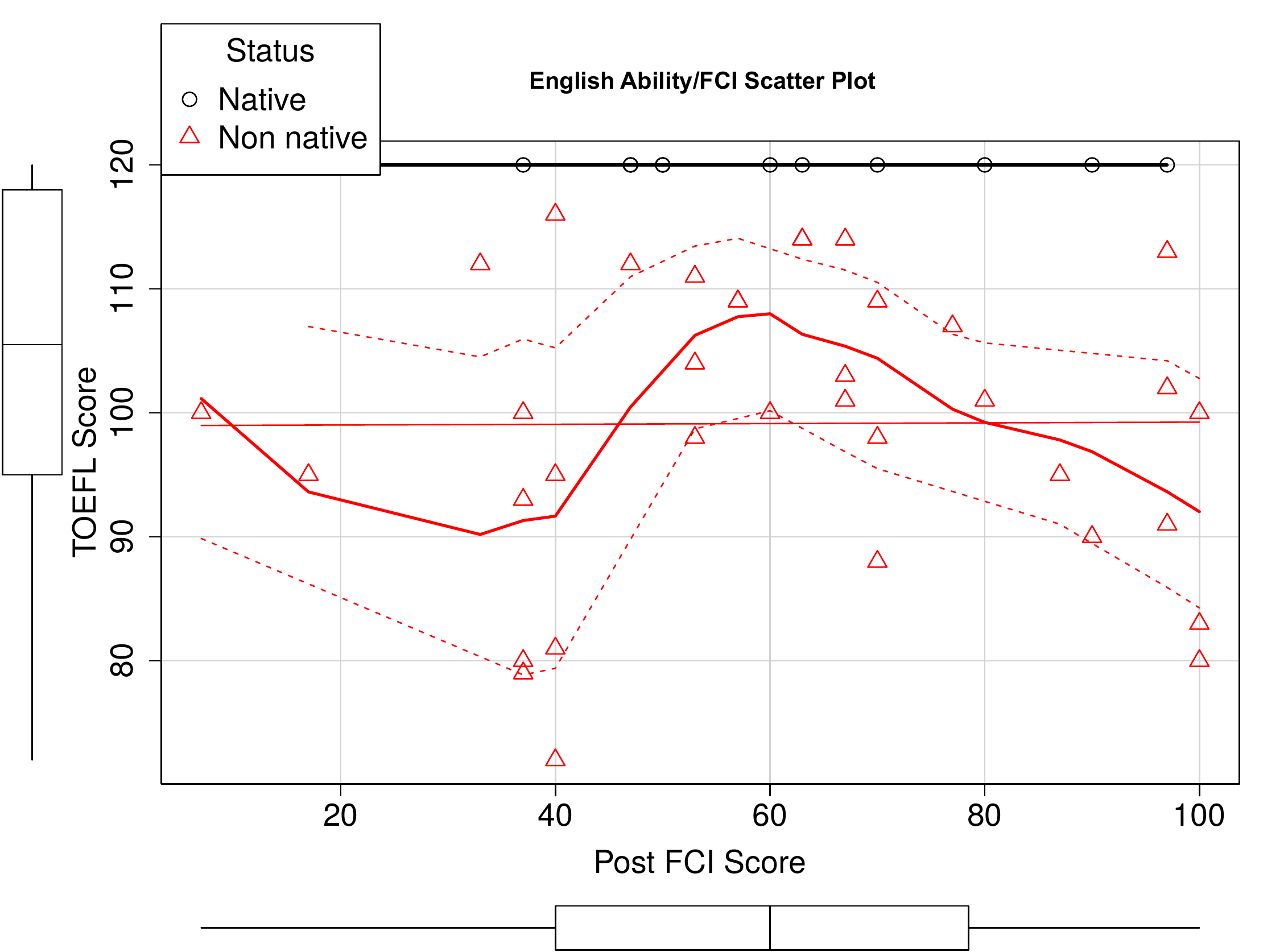}}
    \scalebox{0.6}{\includegraphics{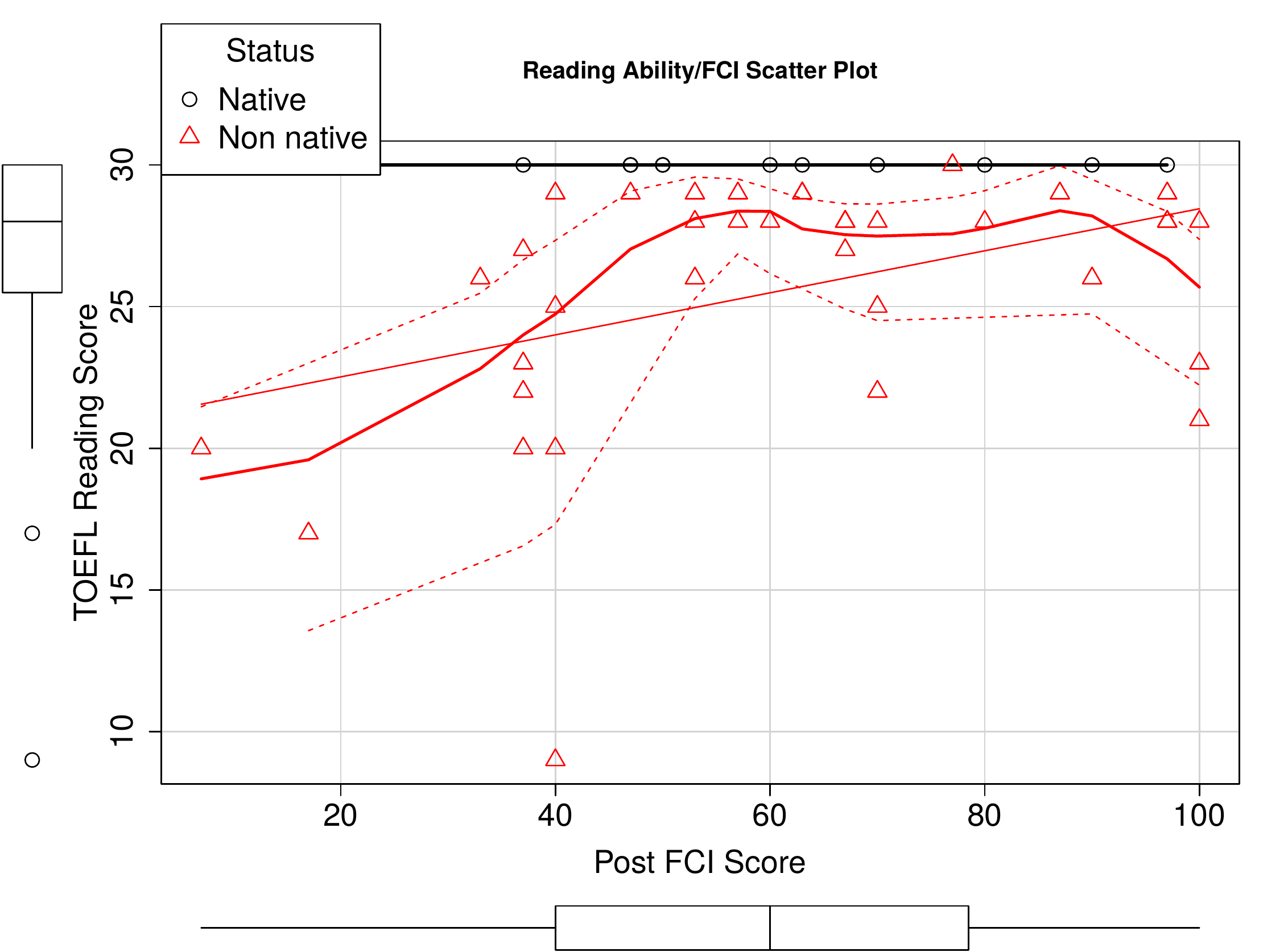}}
    \caption{\it Scatter plot of English (top) \&  reading ability (bottom) vs posttest FCI. We clearly see two groups: Natives and non natives and although we found no correlation between the overall TOEFL and the post FCI score, there appears to be one for non-natives (similar correlations were found for pretest FCI).}
    \label{fig_0}
\end{figure}

 Given that the course is predominantly for non-native speakers of English (the CBCMP entrance requirement is a TOEFL score above $\sim 80$) we considered the possibility that English ability leads to a low $G$. We checked to find a very low correlation between pre/post test performance on the FCI and English ability ($r^2=0.0048/0.00041$, respectively), see  \textbf{Fig. \ref{fig_0}}, including students classed as native. 

\par It may be worth mentioning that the distinction between native and non-native here is blurred because, for administrative reasons a TOEFL/IELTS type score was required for some students who might be considered native (for example even though their language of instruction at high school was English). Actually it is well known that natives do not score full marks on TOEFL/IELTS tests \cite{Stricker}. However, rather than use some kind of complicated probabilistic Monte-Carlo type simulation \cite{Robert} to randomly model ``native" scores for those with TOEFL scores greater than $115$ out $120$, for simplicity we just give full marks to those classed as ``native" administratively. This also applies to reading ability, where in some cases a student is exceptionally better (or worse) than the average taken from their total TOEFL/IELTS score.

\par Perhaps not surprisingly we did find a correlation with reading ability on pre/post FCI with 
$r^2=0.076/0.097$, respectively. This is an order of magnitude larger than the total TOEFL score correlation. The scatter plot in \textbf{Fig. \ref{fig_0}} (lower panel) clearly shows some kind of correlation for non-natives in reading ability, with \textcolor{black}{two} distinct groups. We stress here that this pattern is not the same problem that occurs for native speakers in conceptual tests who have confusion between every day and physics usage of words. This can clearly seen from the fact that we have two separate groups in the scatter plots in \textbf{Fig. \ref{fig_0}}.

\par There are then two possible reasons for the fact that the overall score had no correlation. Either, this is because we are taking a small high-end sample of TOEFL in the range $80 - 120$, while the entrance requirement did not require a minimum reading score on TOEFL; or that reading ability is the more important factor in conceptual tests on global courses. In Japan most institutions only ask for an overall score on English tests, while the evidence here suggests that lower bounds on `reading' and `writing' ability are very important for students to succeed on university courses in general (we hope to discuss further evidence for this in future work).

\end{document}